\documentclass[a4paper,fleqn]{cas-sc}

\usepackage[numbers,sort&compress]{natbib}

\usepackage{bm}

\graphicspath{ {./Figures/} }

\def\tsc#1{\csdef{#1}{\textsc{\lowercase{#1}}\xspace}}
\tsc{WGM}
\tsc{QE}
\tsc{EP}
\tsc{PMS}
\tsc{BEC}
\tsc{DE}

\begin{document}
\let\WriteBookmarks\relax
\def\floatpagepagefraction{1}
\def\textpagefraction{.001}
\shorttitle{FEM Reconstruction of Incongruent Ni–Mn–Ga Single-Crystal Growth}
\shortauthors{D. Musiienko et~al.}

\title [mode = title]{Finite-element reconstruction of molten-zone dynamics and chemical segregation in optical floating-zone grown Ni--Mn--Ga}                      

\author[1]{Denys Musiienko}[type=editor,
                        orcid=0000-0003-0219-3611]
\cormark[1]
\ead{musiienko@fzu.cz}

\credit{Methodology, Validation, Formal analysis, Investigation, Data Curation, Writing - Original Draft, Visualisation}

\affiliation[1]{organization={Department of Magnetic Measurements and Materials, FZU - Institute of Physics of the Czech Academy of Sciences},
            addressline={Na Slovance 1999/2}, 
            city={Prague},
            postcode={182 00}, 
            country={Czech Republic}}

\author[2]{Taras Kovaliuk}[orcid=0000-0002-7712-6758]

\credit{Investigation, Writing - Review \& Editing}

\affiliation[2]{organization={Department of Condensed Matter Physics, Faculty of Mathematics and Physics, Charles University},
            addressline={Ke Karlovu 5}, 
            city={Prague},
            postcode={121 16}, 
            country={Czech Republic}}

\author[3]{Petr Harcuba}[orcid=0000-0002-9910-9158]

\credit{Formal analysis, Investigation, Writing - Review \& Editing}

\affiliation[3]{organization={Department of Physics of Materials, Faculty of Mathematics and Physics, Charles University},
            addressline={Ke Karlovu 5}, 
            city={Prague},
            postcode={121 16},
            country={Czech Republic}}

\author[1]{Oleg Heczko}[orcid=0000-0002-7647-3141]

\credit{Resources, Writing - Review \& Editing, Supervision, Funding acquisition}

\author[2]{Ross H. Colman}[type=editor,
                        orcid=0000-0002-5039-9764]

\credit{Conceptualisation, Methodology, Validation, Formal analysis, Investigation, Resources, Data Curation, Writing - Original Draft, Visualisation, Supervision, Funding acquisition}

\cormark[1]
\ead{ross.colman@matfyz.cuni.cz}

\cortext[cor1]{Corresponding author(s)}

\begin{abstract}
Homogeneous single-crystal growth of Ni--Mn--Ga magnetic shape-memory alloys is limited by incongruent melting, Mn reactivity, and longitudinal composition segregation. We report a modified optical floating-zone method in which growth is performed inside a transparent sapphire crucible, stabilising the molten zone while preserving optical heating and visual monitoring. The resulting crystals show improved growth reproducibility and controlled macroscopic geometry. To interpret the measured segregation profiles, a transient finite-element heat-transfer model in COMSOL Multiphysics was coupled to an effective molten-zone mass-balance calculation. The model reveals that the local solidification velocity is not fixed by the imposed translation speed but varies strongly as the molten-zone geometry evolves. The calculated solidification-front geometry is independently supported by EBSD analysis of a quenched growth interface, while the coupled mass-balance calculation reproduces the dominant features of the measured composition profiles. These results show that deviations from ideal Scheil--Pfann behaviour arise from the evolving molten-zone geometry and solidification conditions. The experimentally constrained model therefore provides a practical means of reconstructing otherwise inaccessible growth conditions and a basis for optimisation of sapphire-confined floating-zone growth of incongruently melting alloys.
\end{abstract}

\begin{graphicalabstract}
\includegraphics{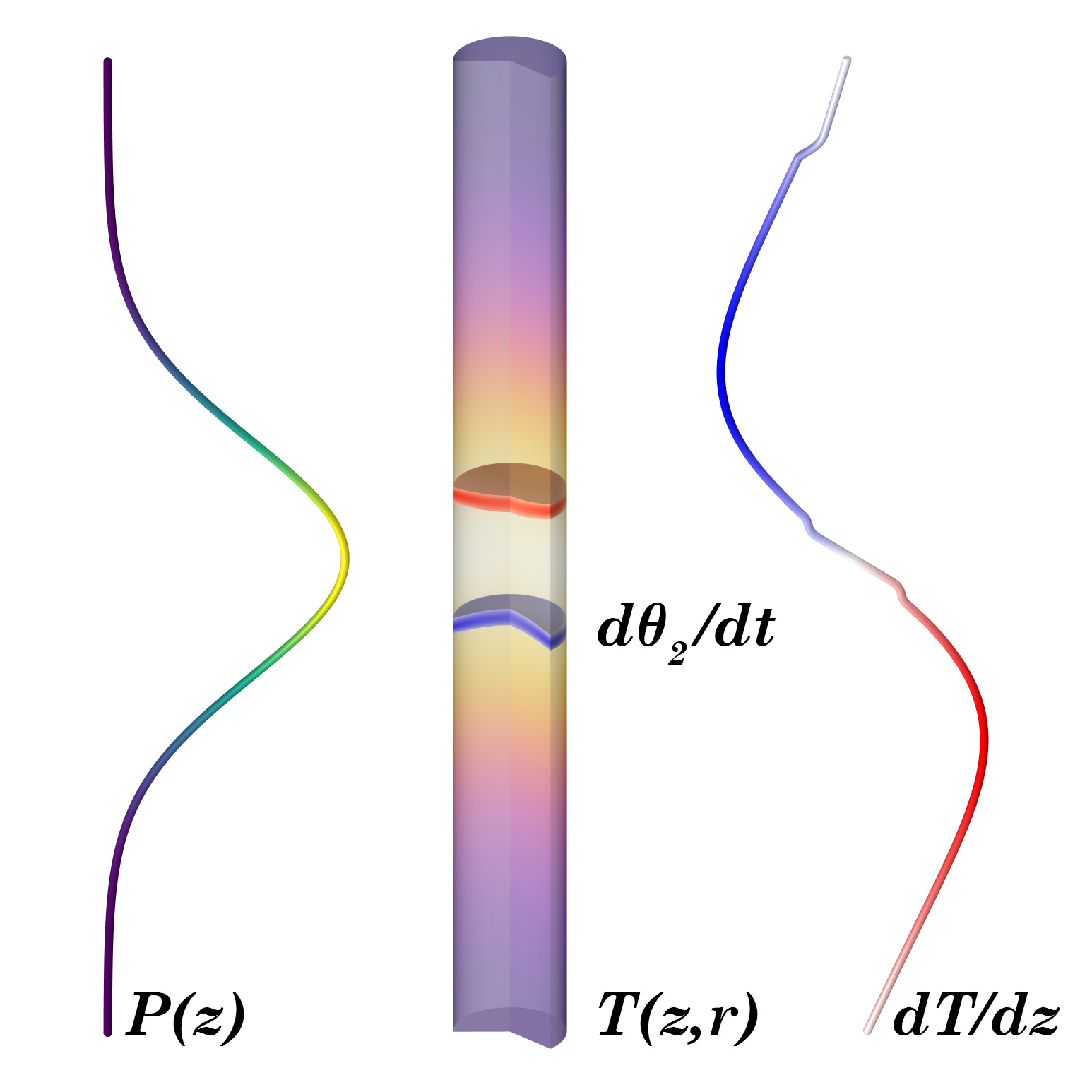}
\end{graphicalabstract}

\begin{highlights}
\item Sapphire confinement enables stable optical floating-zone growth of Ni–Mn–Ga.
\item Coupled finite element method and mass-balance modelling reproduces the molten-zone evolution and composition segregation.
\item Electron backscatter diffraction of a quenched sample confirms the convex solidification-front geometry predicted by the model.
\item Local solidification velocity differs substantially from the imposed translation rate.
\end{highlights}


\maketitle

\section{Introduction}
\label{sec:Introduction}
Incongruently melting materials, where the composition of the solid differs from that of the melt during solidification, present a challenge for crystal growers. Floating zone techniques where only a portion of the charge volume is molten at a given moment of the growth process are known to help mitigate the effects, but macroscopic chemical segregation - concentration gradients - are still typically observed. This problem is particularly important in functional alloys such as Ni--Mn--Ga, where relatively small composition variations can strongly affect the phase transformation temperatures and resulting functional properties.

Magnetic shape memory (MSM) smart materials are of significant interest for technological applications. Since their first experimental demonstration in 1996~\cite{ullakko1996large}, the Ni--Mn--Ga Heusler alloys have been extensively studied~\cite{o2000phenomenology, likhachev2000magnetic, Sozinov2002, Heczko2009, straka2011highly, Kohl2014, saren2017dynamic, takhsha2025spatially}. The initially observed strains of 0.2\% were rapidly improved upon, with the highest observed strain currently at 12\%~\cite{sozinov201312}. They are being investigated for industrial applications ranging from microfluidic pumps~\cite{smith2015characterization}, vibrational energy harvesters~\cite{saren2015modeling, Farsangi2017}, and actuators~\cite{Holz2012}, to incorporation in kinetic energy absorbing polymers~\cite{Lahelin2009, Feuchtwanger2005}. Whilst several polycrystalline forms, including wires~\cite{Wang2015}, printed structures and foams are being trialled~\cite{chmielus2009giant, Witherspoon2015, laitinen2024powder}, single crystalline Ni--Mn--Ga material is required to maximise the MSM effect~\cite{laitinen2022giant}.

Moreover, the stoichiometric alloy exhibits MSM effect only below 200 K (its martensite transformation temperature), and to increase the working temperature a composition modified from the ideal Heusler 2-1-1 stoichiometry is needed. Minor composition changes result in dramatic shifts of the onset of the martensitic transformation, $\sim$ 28 K/at.\%$_{\text{Mn}}$ in the Ni$_{50}$Mn$_{25+x}$Ga$_{25-x}$ system~\cite{jin2002empirical}. The ability to easily tune the structural transition is an important feature of the Ni--Mn--Ga family. Investigations of substitutional doping with other transition metals has further expanded the function of these alloys by increasing the operating temperature range~\cite{perez2019study, armstrong2020systematic, kowalska2025transformation} and increasing twin boundary mobility~\cite{sozinov2020highly, perez2020role, ding2023strain}. 

Bulk Ni--Mn--Ga single crystals have been grown using several directional-solidification approaches. Early studies employed the conventional Bridgman method, demonstrating bulk single-crystal growth but also revealing pronounced longitudinal chemical segregation associated with the incongruent solidification of Ni--Mn--Ga~\cite{schlagel2000chemical}. Modified Bridgman and Bridgman--Stockbarger (BS) techniques, including seeded growth, were subsequently developed to improve crystallographic orientation and growth stability and have been used to produce high-quality crystals for magnetic-shape-memory studies~\cite{brillo2011thermophysical,Kellis2012}. Nevertheless, such compositional variation, while useful for scientific studies spanning a broad range of Ni--Mn--Ga compositions, is undesirable for technological applications, where reproducible functional behaviour requires tight compositional control~\cite{Fan2021}. Zone-melting methods offer a means of reducing these longitudinal composition gradients by restricting the volume of molten material and allowing the melt composition to approach a quasi-steady segregation state during translation~\cite{Pamplin1980,Jiang2005,Okamoto2006}. Optical floating-zone (OFZ) growth was explored as a crucible-free implementation of this approach, avoiding direct contact between the molten alloy and a conventional crucible while providing control over the molten-zone geometry, although segregation and the stability and shape of the solid--liquid interfaces remain important limitations~\cite{Wang2008,Liu2012,Liu2011,cejpek2019rapid}. A stable and well-defined solidification front for OFZ is particularly important because its geometry and local velocity govern the direction of crystal growth, grain selection, and the development of defects and sub-grain boundaries.

In the present work, we present the results of growth experiments in a sapphire-confined modified optical floating-zone geometry, overcoming several limitations of the standard optical floating-zone growth. We complement this modified growth method with a time-dependent finite element heat-transfer model, coupled with an effective molten-zone mass-balance calculation. Using the measured post-growth composition profile together with the initial compositions of the seed crystal and feed rod, the model reconstructs the evolving molten-zone shape and local solidification velocity during growth, rather than assuming a constant molten-zone length and equating solidification velocity with the imposed crucible translation rate, as in classic Scheil--Pfann descriptions.

\section{Experimental methods}
\label{sec:ExpMethods}
\subsection{Alloy preparation}
\label{subsec:AlloyPrep}
The elemental metals Ni (99.995\%), Ga (99.999\%) and Mn (99.98\%, etched prior to use by dilute HNO$_{3(\text{aq})}$) all purchased from Alfa Aesar, were weighed to give a nominal alloy composition of Ni$_{50}$Mn$_{28.5}$Ga$_{21.5}$. Pre-alloying was performed by inductive heating in an alumina crucible within an RF furnace operating at 100~kHz under argon atmosphere. The power was ramped to 4.5~kW over 5 min, followed by a 30~s hold to promote mixing before drop-quenching into a copper mould of 6~mm diameter and 80~mm length. This resulted in high-density polycrystalline ingots with homogeneous composition along the length. 

\subsection{Sapphire-confined optical floating-zone growth}
\label{subsec:OFZgrowthMethod}

Stable crucible-free OFZ growth of Ni--Mn--Ga was difficult to maintain because of oxide-crust formation at the melt surface. This behaviour is attributed to the high reactivity and vapour pressure of Mn near the melting temperature, producing a Mn-rich vapour that reacts with trace residual oxides released from furnace surfaces. The resulting surface crust remains stable even under a 5\% H$_2$ reducing atmosphere~\cite{Eijk2023} and can obscure the molten-zone geometry, and disturb heat transfer. Such disturbances can destabilise the molten zone and ultimately interrupt the growth process~\cite{Li2008}. To mitigate these effects, the conventional floating-zone geometry was modified by introducing a transparent single-crystal sapphire crucible.

\begin{figure}[pos=htbp]
    \centering
    \includegraphics[width=0.45\linewidth]{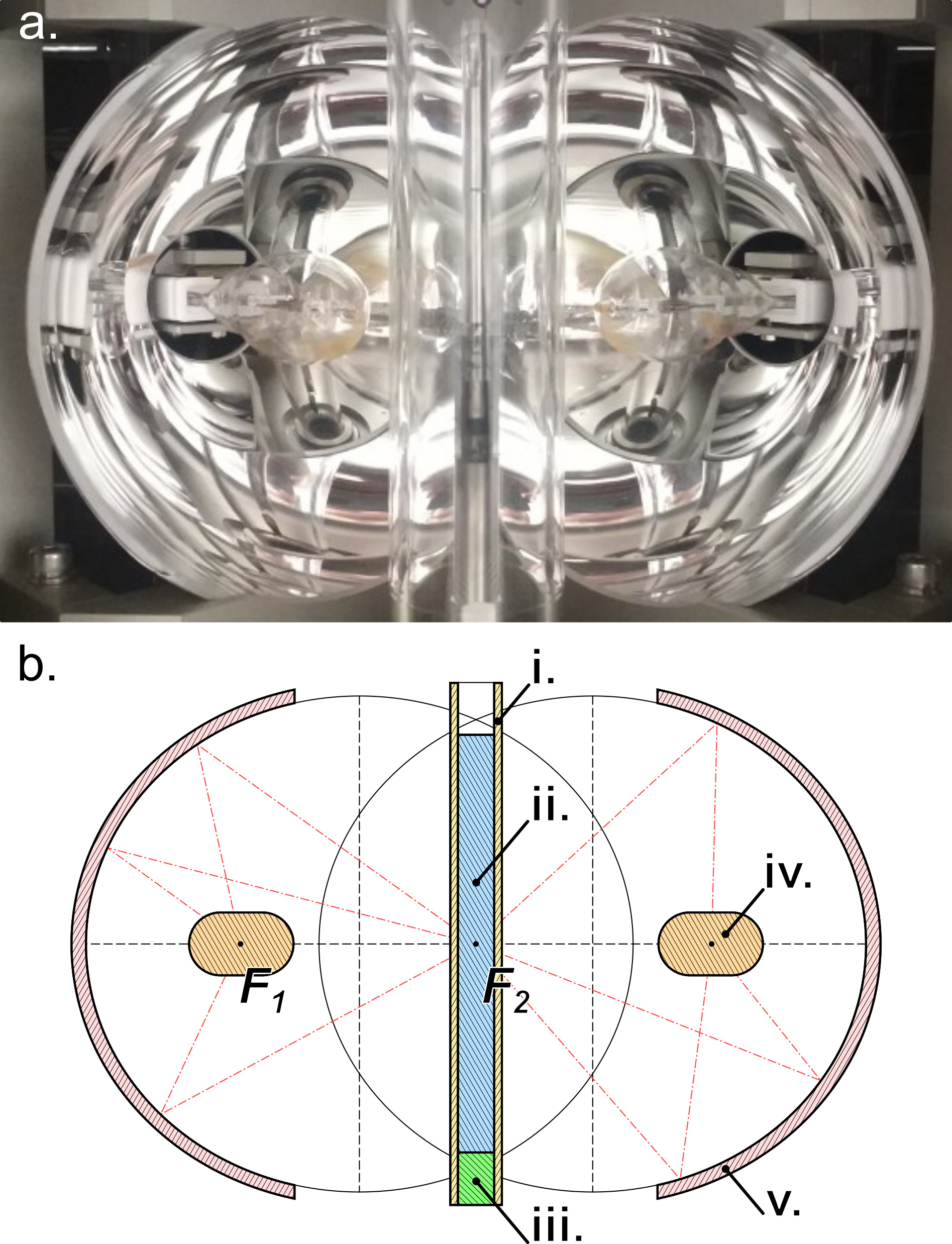}
    \caption{{\bf a.} Photograph, and {\bf b.} schematic of in-crucible growth, utilizing a 4-mirror optical floating zone furnace from Crystal Systems Corp. With components: i. sapphire crucible; ii. precursor polycrystal charge; iii. Oriented seed; iv. halogen lamps; v. elliptical mirrors with focal points F1 and F2.}
    \label{fig:OFZschematic}
\end{figure}

A photograph of the experimental setup and a simplified schematic are shown in Fig.~\ref{fig:OFZschematic}. Crystal growth was performed using a four-mirror optical floating-zone furnace (Crystal Systems Corp. FZ-T-4000-VI-VPM-PC) equipped with 1000 W halogen lamps and modified for high-vacuum operation. An oriented single-crystal seed, 10 mm long and aligned with the austenite [100] direction along the growth axis, was placed at the bottom of the sapphire tube (7 mm diameter, 150 mm length; manufactured by CRYTUR, spol. s r.o., Turnov, Czech Republic\footnote{https://www.crytur.com/solutions/sapphire-precision-optics/}), with the polycrystalline feed rod positioned above it. The seed and feed-rod diameters were slightly smaller than the inner diameter of the tube to accommodate different thermal expansions of sapphire and Ni--Mn--Ga alloy during heating. Before each growth, the furnace was evacuated to below $10^{-5}$ mbar and back-filled with Ar (99.9999\%) to an overpressure of 2.4 bar. The lower end of the feed rod, in contact with the seed crystal, was melted by slowly increasing the furnace power over approximately 3 h until a stable molten zone of $\approx10$ mm height was visually established, typically at 42 \% of the furnace power. Growth was then carried out without rotation at a translation rate of 5 mm/h over the full length of the ingot.

\subsection{Composition and microstructural characterisation}
\label{subsec:CompMicrStructCharact}
Samples for all characterisation methods were prepared by mechanical grinding and polishing, followed by chemical etching using a copper-sulfate based nickel-alloy etchant, known in the laboratory as \textit{blue poison}. The etchant consisted of 3.5~ml H$_2$SO$_4$ (98\%), 50~ml HNO$_3$ (65\%), 15~g CuSO$_4$, and 15~ml H$_2$O. Composition analysis was performed by X-ray fluorescence spectroscopy EDAX AMETEK ORBIS PC, equipped with a Rh anode tube and Apollo XRF ML-50 EDS detector. The incident beam was defined by the instrument’s nominal 50 µm microcapillary optics, and the measurements were performed under dynamic vacuum. Electron backscatter diffraction (EBSD) was used to determine crystal orientation and to map the solidification-front morphology in crystals quenched during growth by switching off the furnace lamps before complete re-melting of the feed rod. Measurements were performed using a Zeiss Auriga Compact scanning electron microscope equipped with a Velocity EBSD detector (EDAX AMETEK) and in-situ heating stage. EBSD data were acquired at an accelerating voltage of 20~kV and subsequently analysed using OIM Analysis software (EDAX AMETEK).

\section{Model development and experimental validation}
\label{sec:ModelMethods}

\subsection{Composition segregation problem}
\label{subsec:CompSegr}
\begin{figure}[pos=htbp]
    \centering
    \includegraphics[width=0.5\linewidth]{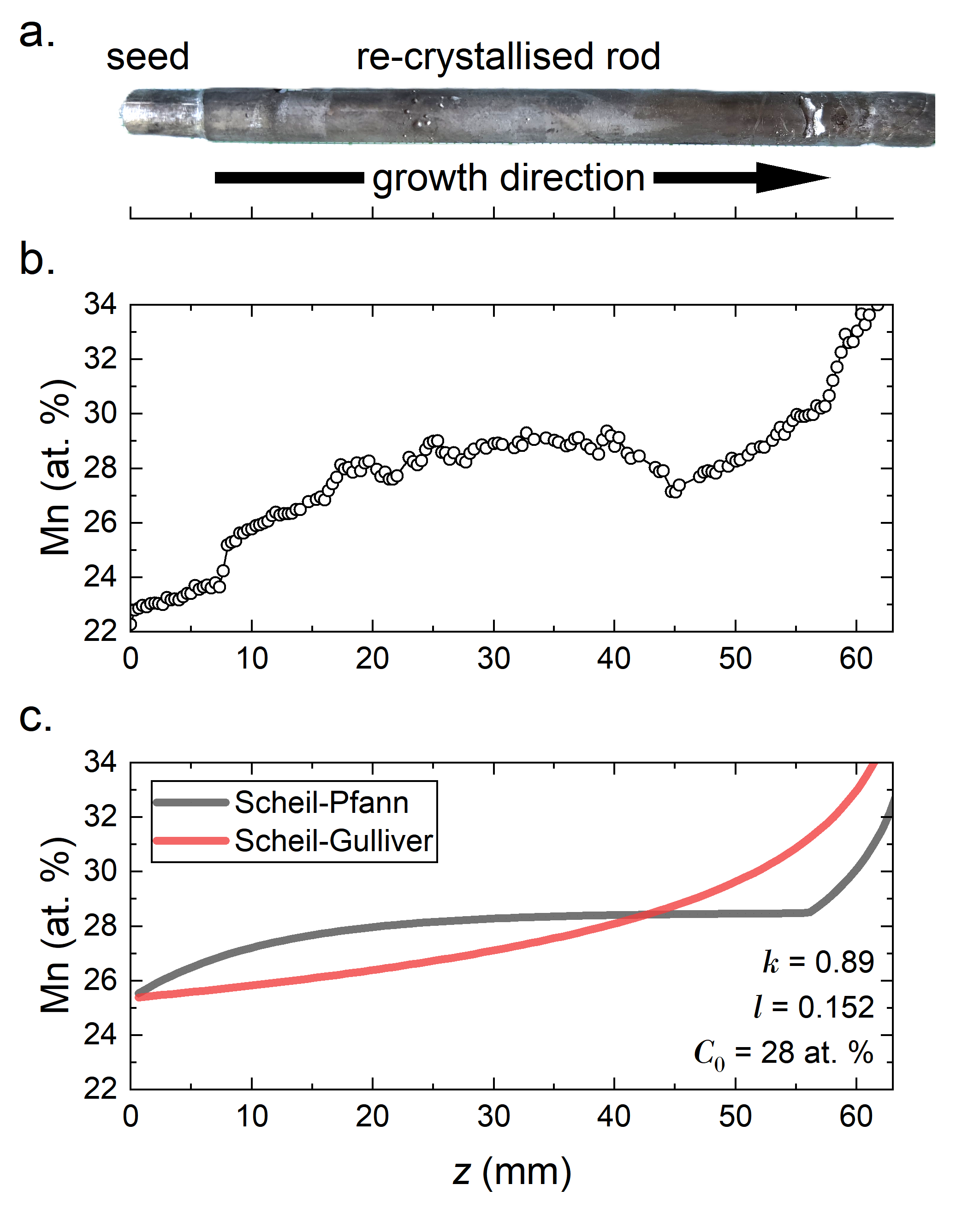}
    \caption{Composition variation, as atomic percent Mn in floating zone growth of Ni--Mn--Ga alloy. {\bf a.} photo of resulting ingot from an in-crucible growth; {\bf b.} measured composition variation along the growth direction; and {\bf e.} calculated macroscopic segregation expected in floating zone (Scheil--Pfann) and Bridgman (Scheil--Gulliver) growth types, discussed in the text.}
    \label{fig:intial_growths}
\end{figure}

The use of a sapphire crucible results in a controlled melt diameter during the growth, see Fig.~\ref{fig:intial_growths}a, however growth of homogeneous composition Ni--Mn--Ga crystals of useful length is plagued by a further complication: longitudinal chemical segregation due to the incongruently melting characteristics of the Ni--Mn--Ga system, see Mn variation along the growth direction in Fig.~\ref{fig:intial_growths}b. At the growth interface, it is more energetically favourable to crystallise with a reduced Mn concentration and an enriched Ga and Ni concentration than is present in the melt. When using a growth technique that relies on complete melting of the entire charge, such as in BS growth methods, and when the rate of crystallisation is slow enough to allow complete mixing in the melt, this leads to a continuously changing  melt composition (and therefore a continuously changing solid composition) along the growth length. The governing equations describing this behaviour, are known as the Scheil--Gulliver equations~\cite{OSTROGORSKY2015995}:
\begin{equation}
    C(f) = kC_0(1-f)^{k-1}
    \label{eq:SheillGulliver}
\end{equation}
where, $C(f)$ and $C_0$ are the current and initial compositions at solidification fraction $f$, for each element with partition coefficient $k$. This continuously evolving behaviour has been observed previously in Ni--Mn--Ga single crystals produced by BS growth method~\cite{schlagel2000chemical}, and was used to determine the partition coefficients of $K_{Ni} = 1.04$, $K_{Mn} = 0.89$, and $K_{Ga} = 1.06$.

A partial melting of the charge, such as in zone melting, reduces these segregation effects and results in a modification of the Scheil--Gulliver equations, known as the Scheil--Pfann equation~\cite{OSTROGORSKY2015995}:  
\begin{equation}
C(f) = C_0\cdot
    \begin{cases}
    1-(1-k)e^{-k\frac{f}{l}}, & 0 \le f \le 1 - l \\[6pt]
    \left[1-(1-k)e^{-k\frac{1-l}{l}}\right]\cdot\left(\frac{1-f}{l}\right)^{k-1}, & 1 - l < f \le 1
    \end{cases}
    \label{eq:ScheilPfann}
\end{equation}
where $l$ is the molten-zone length fraction. The resulting expected behaviour for both the Scheil--Gulliver (BS), and the Scheil--Pfann (OFZ) equations are shown in Fig.~\ref{fig:intial_growths}c. However, they are predicated on several assumptions, including a slow-growth speed that allows complete mixing of the melt, and maintaining a constant volume of the molten zone. It is clear from Fig.~\ref{fig:intial_growths} that whilst the Scheil--Pfann equation captures the broad behaviour of the chemical segregation during our modified OFZ growth, subtle breaking of these assumptions prevents the classic model from perfectly matching the experimentally observed segregation. 

Classical Scheil--Pfann description requires a prescribed solidification geometry and velocity, whereas in the sapphire-confined OFZ configuration the actual molten-zone size and its evolution during growth cannot be reliably monitored due to wetting of the sapphire by the melt. A time-dependent heat transfer model is therefore required to reconstruct the molten-zone geometry and local solidification conditions before the segregation behaviour can be interpreted quantitatively.

\subsection{Finite-element heat-transfer model}
\label{subsec:FEHTmodel}
A transient heat-transfer model of the optical floating-zone experiment was developed in COMSOL Multiphysics\textsuperscript{\textregistered} 6.2 using an axisymmetric geometry reproducing the experimental configuration. Heat transport within the Ni--Mn--Ga rod was described by conduction, with radiative and convective heat losses applied at the external boundaries. Optical heating was represented by a Gaussian heat-flux profile. To account for translation of the heating zone, the profile was evaluated in a coordinate system moving with the focal point, following the conditional boundary implementation described by S{\"o}nnerlind~\cite{Sonnerlind2016ConditionalBC}. Its width parameter was fixed at $\omega=0.016$~m ($\omega=2\sigma$), with $\mathrm{FWHM} = \sqrt{2\ln{2}}\:\omega\approx 0.0188\:\mathrm{m}$ corresponding to the effective filament length perpendicular to the growth direction. Thermophysical properties of the Ni--Mn--Ga alloy were taken from Brillo \textit{et al.}~\cite{brillo2011thermophysical}. Constant model parameters are summarised in Table~\ref{tab:heattransfer}, while the temperature dependences of heat capacity and thermal conductivity were implemented as interval functions, and are defined in Appendix~\ref{app:A}.

\begin{table}[ht]
    \caption{Physical parameters used in the Heat Transfer model.}
    \label{tab:heattransfer}    
    \centering
    \begin{tabular*}{\tblwidth}{@{}LLL@{}}
        \toprule
        Parameter & Value & Description \\
        \midrule
        $P_{\mathrm{integral}}$ & $65\;\mathrm{(W)}$ & Deposited heating power\\
        $v_{\mathrm{vert}}$ & $5\;(\mathrm{mm\cdot h}^{-1})$ & Vertical translation speed\\
        $T_{m}$ & $1393\;\mathrm{(K)}$ & Melting point\\
        $\Delta T_{m}$ & $20\;\mathrm{(K)}$ & Temperature interval of phase transformation\\
        $L$ & $223\;\mathrm{(kJ\cdot kg^{-1})}$ & Latent heat of melting/solidification\\
        $\rho$ & $6900\;\mathrm{(kg\cdot m^{-3})}$ & Material density near phase transformation\\
        $h_{\mathrm{conv}}$ & $20\;(\mathrm{W}\cdot\mathrm{m}^{-2}\cdot\mathrm{K^{-1}})$ & Effective convective heat-transfer coefficient\\
        $\epsilon$ & $0.2\;(1)$& Surface emissivity\\
        $T_{\mathrm{amb}}$ & $300\;\mathrm{(K)}$ & Ambient temperature\\
        \bottomrule
    \end{tabular*}
\end{table}

\begin{figure}[pos=htbp]
    \centering
    \includegraphics[width=1.0\linewidth]{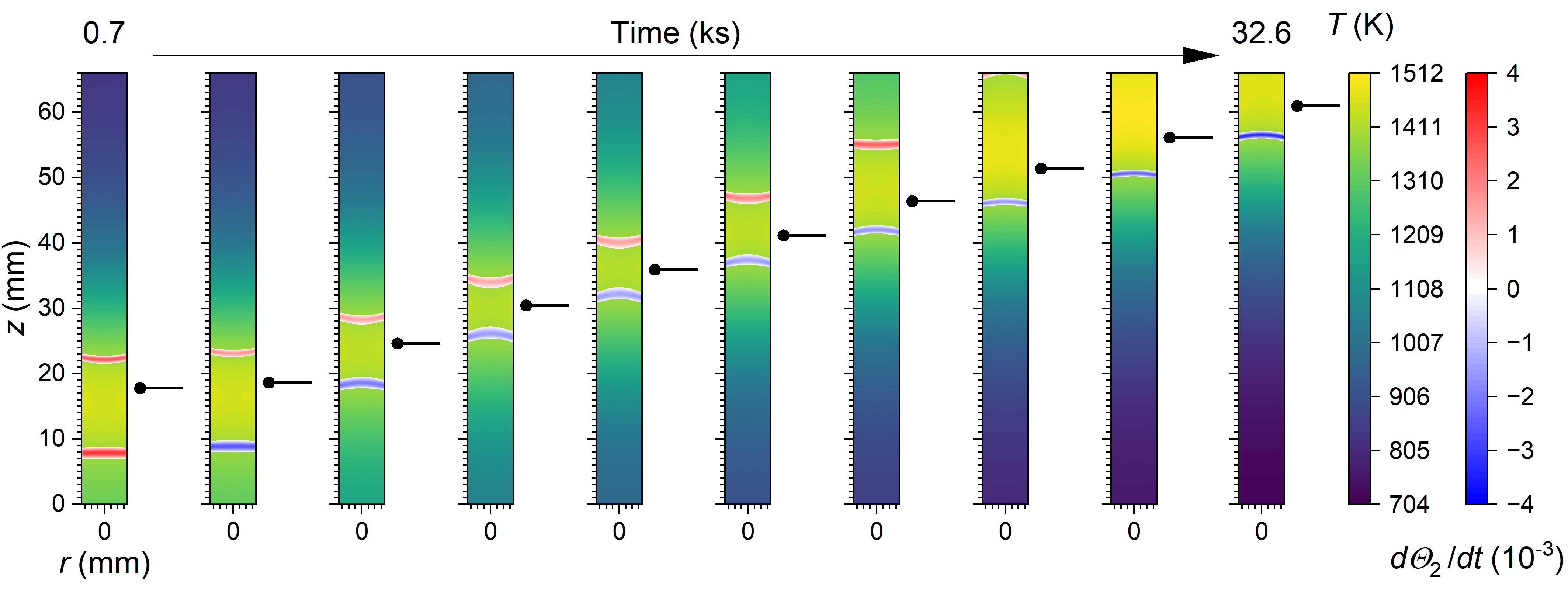}
    \caption{Simulation snapshots at selected time steps, showing the ingot cross-sectional temperature profile, phase change interfaces, and the focal position of the lamps, $z_f$, dot-terminated black line.}
    \label{fig:HTModelDemo}
\end{figure}

Melting and solidification were treated using an apparent-heat-capacity formulation incorporating the latent heat over a finite temperature interval around the melting point. Radiative losses were calculated using an effective emissivity of $\varepsilon = 0.2$, while the convective heat-transfer coefficient was fixed at $h_{\mathrm{conv}} =20\;\mathrm{W}\mathrm{m}^{-2}\mathrm{K^{-1}}$. The resulting transient temperature field was used to determine the solid-liquid interfaces, molten-zone height, and local solidification velocity throughout the simulated growth. Full governing equations, boundary conditions, numerical implementation, and parametrisation are provided in Appendix~\ref{app:A}.

The calculated temperature field provides direct access to the evolution of the molten-zone geometry during growth. In contrast to the classical Scheil--Pfann description, which assumes a constant molten-zone volume, the present model predicts significant variation in molten-zone size throughout the process, as illustrated by the selected time snapshots in Fig.~\ref{fig:HTModelDemo}. This variation arises naturally from the evolving thermal balance in the finite-length rod geometry.

During the first half of the growth, the molten-zone volume gradually decreases. This is attributed to the increasing length of solidified crystal below the molten zone, which enhances heat extraction from the melt. As the molten zone approaches the middle of the ingot, this effect becomes progressively balanced by the shortening of the feed rod above the molten zone, which reduces the heat-sinking capability on the upper side. Beyond the midpoint, the continuing reduction of material above the molten zone dominates, and the molten-zone volume increases again until the remaining feed rod is fully molten.

This thermal asymmetry is also reflected in the axial temperature distribution. As shown in Fig.~\ref{fig:HTModelDemo}, the molten zone does not remain centred on the optical focal position, $z_f$, but becomes increasingly shifted towards the side with the lower remaining thermal mass. Thus, the computed temperature field reveals not only changes in molten-zone volume, but also a progressive offset between the molten zone and the nominal heating focus. In longer growths, such end effects would be less pronounced once the molten zone is sufficiently far from both ends; however, they are unavoidable in the present laboratory-scale geometry, where the molten zone represents a substantial fraction of the total crystal length.

\subsection{Coupled molten-zone mass-balance model}
\label{subsec:MassBalanceModel}

The transient thermal model was coupled to an effective molten-zone mass-balance calculation to reconstruct the compositional evolution during re-solidification. Mn and Ga were tracked explicitly, as they show the largest relative segregation in the studied Ni--Mn--Ga composition range, while Ni was calculated as the remaining fraction. The initial solid composition was prescribed from the experimentally measured longitudinal composition profiles of the seed crystal and polycrystalline feed rod.

\begin{figure}[pos=htbp]
    \centering
    \includegraphics[width=0.5\linewidth]{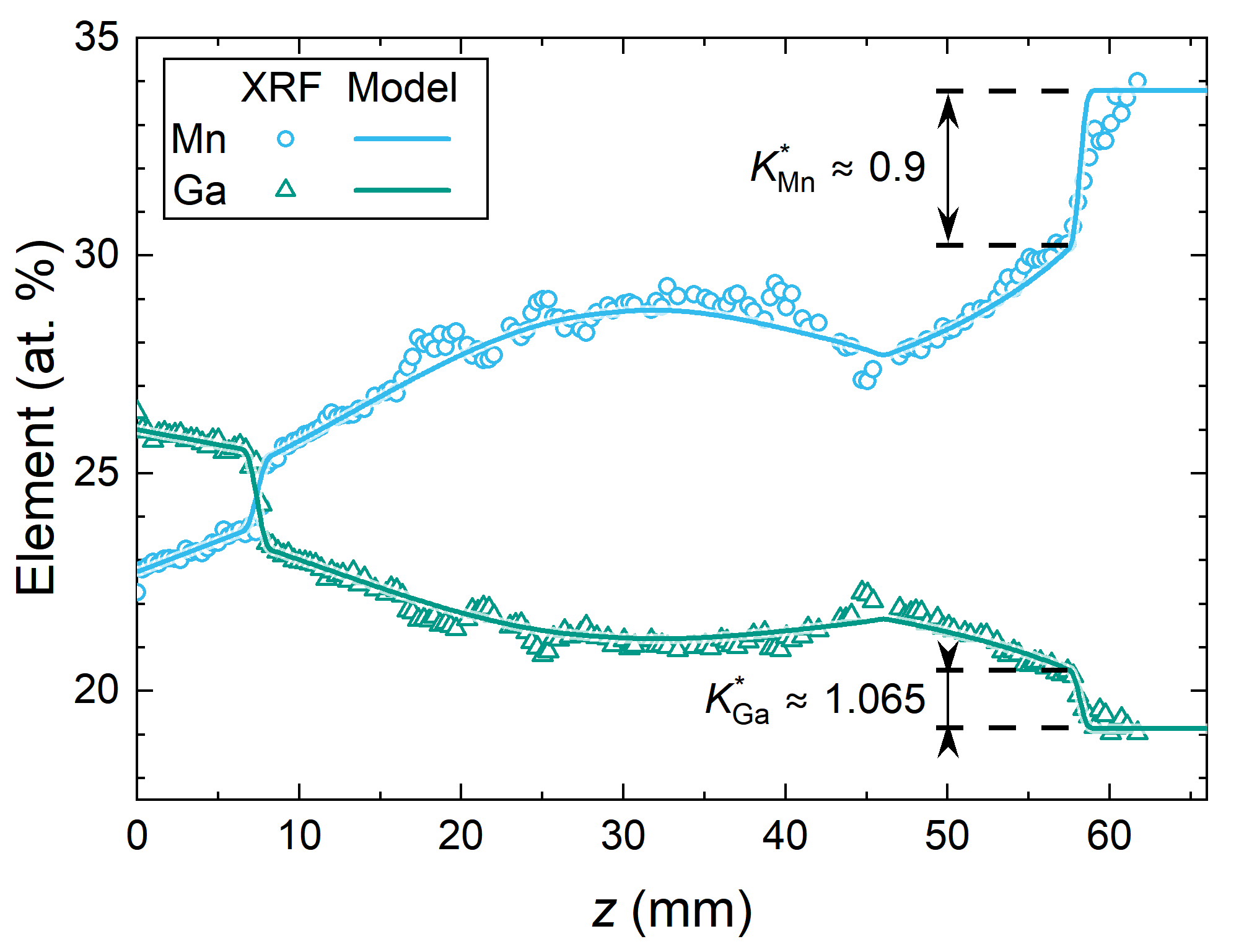}
    \caption{Measured (open symbols) and calculated (solid lines) elemental composition profiles along the crystal growth direction, $z$. The effective partition coefficients used in the model are denoted by $K^*_{\mathrm{Mn}}$ and $K^*_{\mathrm{Ga}}$.}
    \label{fig:XRFvsModelConc}
\end{figure}

At each time step, the phase fractions obtained from the heat-transfer calculation were used to determine the volumes of material entering and leaving the molten zone through melting and solidification. The melt was treated as compositionally homogeneous, with newly melted material incorporated into a single molten-zone composition. Material leaving the melt was assigned a solid composition using effective partition coefficients, $K^*_{\mathrm{Mn}}$ and $K^*_{\mathrm{Ga}}$, after which the molten-zone composition was updated by conservation of each tracked element. In this way, the changing molten-zone volume and interface motion calculated by the thermal model directly determine the evolution of both the liquid and re-solidified compositions.

The effective partition coefficients were constrained experimentally from the sharp composition change observed near the end of the re-solidified bar, using the mean compositions measured immediately before and after this transition, as indicated in Fig.~\ref{fig:XRFvsModelConc}. They therefore describe the effective segregation under the present growth conditions rather than equilibrium partitioning alone. The complete mass-balance formulation, state-variable definitions, time-step bookkeeping, and numerical implementation are given in Appendix~\ref{app:A}.

The resulting experimentally constrained HT--mass-balance model reproduces the dominant features of the measured longitudinal segregation profiles in Fig.~\ref{fig:XRFvsModelConc}. After the initial composition discontinuity associated with melting back into the seed region, Mn becomes progressively enriched and Ga depleted during the first part of the growth. Around the middle of the re-solidified length, these gradients reverse as the molten-zone geometry evolves. Once the remaining feed rod is fully molten, the system no longer behaves as a bounded floating zone and the segregation trend changes again, approaching a Bridgman-like solidification regime in which only the melt and advancing solidification front remain. Finally, switching off the furnace lamps quenches the residual melt, producing the sharp terminal Mn and Ga composition jump towards the instantaneous melt composition. These changes arise from the evolving molten-zone volume and geometry calculated by the thermal model rather than being imposed directly in the segregation calculation.

The effective partition coefficients obtained here are close to those previously derived from Scheil--Gulliver fits to macroscopic segregation in Bridgman-grown Ni--Mn--Ga ingots~\cite{schlagel2000chemical}. Nevertheless, $K^*_{\mathrm{Mn}}$ and $K^*_{\mathrm{Ga}}$ should be regarded as phenomenological parameters incorporating equilibrium partitioning together with deviations arising from finite liquid mixing, local solidification conditions, and the simplified treatment of solute transport. The coupled model therefore provides an intermediate description between conventional analytical segregation models and a fully coupled thermo-fluid and species-transport simulation.

The present experiments were performed at a growth rate of 5~$\mathrm{mm\;h^{-1}}$, compared with 2~$\mathrm{mm\;h^{-1}}$ in the previous BS study \cite{cejpek2019rapid}. The similarity of the fitted effective partition coefficients suggests that the macroscopic segregation behaviour remains comparable within this slow-growth regime and supports their use for simulations at similar growth rates. At higher solidification velocities, reduced time for solute redistribution is expected to decrease the apparent macroscopic segregation and shift effective partition coefficients towards unity. This may reduce longitudinal composition gradients, although faster growth can also promote cellular or dendritic interface instabilities~\cite{Chen2003,cejpek2019rapid}.

\subsection{Solidification interface geometry}
\label{subsec:SolidusValidation}
\begin{figure}[pos=htbp]
    \centering
    \includegraphics[width=0.45\linewidth]{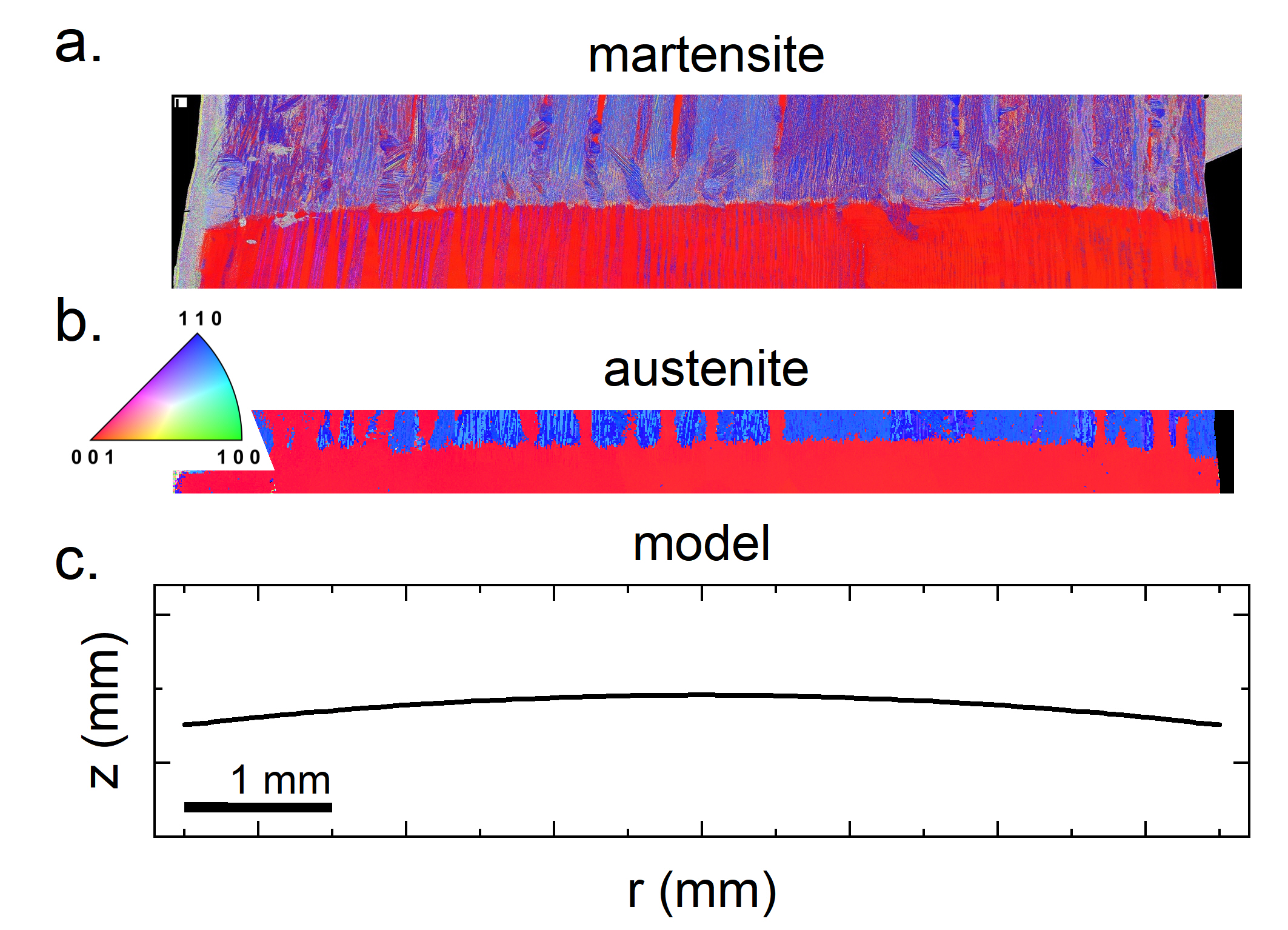}
    \caption{Qualitative comparison between a quenched solidification interface from an independently grown Ni–Mn–Ga ingot, \textbf{a,b.} imaged EBSD (austenite lattice coordinates), and \textbf{c.} calculated solidus-interface profile from the simulation.}
    \label{fig:EBSD}
\end{figure}

To obtain an experimental reference for the solidification-front geometry, one growth was interrupted before complete remelting of the feed rod by switching off the furnace lamps, thereby preserving the approximate position and shape of the interface. EBSD maps acquired in the martensitic and austenitic states, shown in Fig.~\ref{fig:EBSD}a,b, directly map the local crystallographic orientation and confirm the continuity and orientation of the grown crystal. The spatial transition between the re-solidified crystal and the remaining unremelted feed rod delineates the preserved solidification front, which exhibits a convex profile across the ingot radius. Such interface geometry influences the local growth direction and the propagation of competing grains and sub-grain boundaries; a slightly convex front is generally considered favourable for directing such boundaries towards the outer surface of the crystal~\cite{Kobayashi1978}.

The corresponding thermal model independently provides the instantaneous molten-zone geometry and position of the solidification interface. For this particular growth, the deposited heating power and initial focal position were adjusted to reproduce the measured longitudinal composition profile and experimental growth conditions, while the interface shape itself was not fitted. The calculated solidus profile shown in Fig.~\ref{fig:EBSD}c exhibits a convex curvature consistent with the experimentally preserved interface. This agreement provides an additional validation of the calculated temperature field and supports its use to determine the evolving molten-zone height and local solidification velocity during growth.

Approximating the preserved solidification-front profile observed in the martensitic-state EBSD map by a circular arc gives a radius of curvature of $R_{\mathrm{exp}}=31.1$ mm, compared with $R_{\mathrm{mod}}=30.7$ mm for the calculated solidus profile, a relative difference of approximately 1.3 \%. For an interface diameter of 7.0 mm, these radii correspond to centre-to-edge interface heights of approximately 0.198 and 0.200 mm, respectively. The corresponding dimensionless convexities, $h/a$, where $a$ is the crystal radius, are approximately 0.056 and 0.057, quantifying the shallow convex profile towards the melt. This normalized measure facilitates comparison between growths with different crystal diameters ~\cite{Hossain2014}.

\subsection{Molten zone evolution and re-solidification speed}
\label{subsec:MZevolutionResolidSpeed}

\begin{figure}[pos=htbp]
    \centering
    \includegraphics[width=0.45\linewidth]{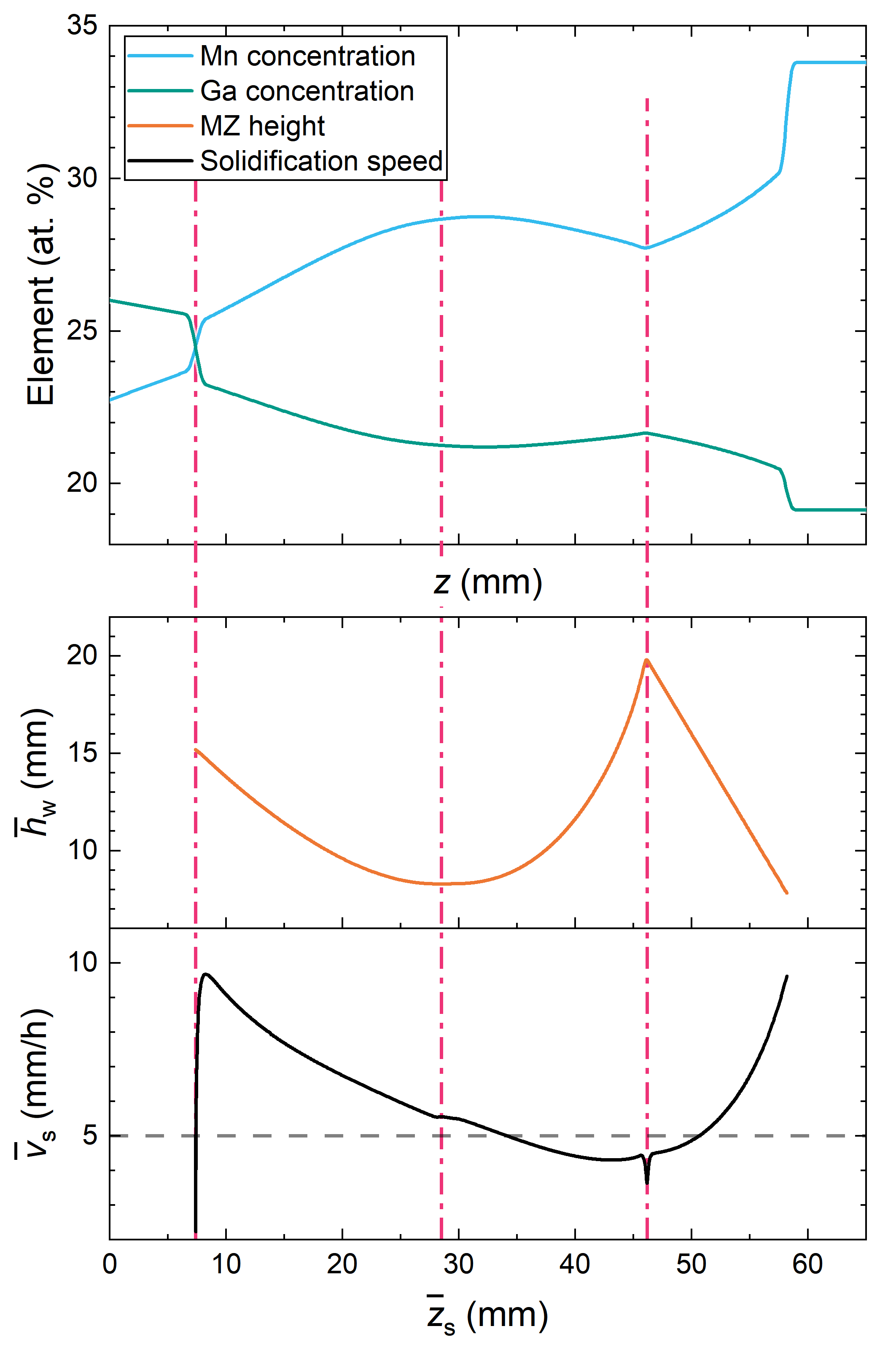}
    \caption{Comparison of calculated elemental composition segregation plotted against the crystal length, $z$, and weighted averages of molten-zone height, $\overline{h}_{\mathrm{w}}$, and solidification-interface velocity, $\overline{v}_{\mathrm{s}}$. molten-zone height (orange) and solidification speed (black) are plotted against the weighted average coordinate of solidification interface, $\overline{z}_s$. Crucible translation speed is denoted by dashed horizontal line. Vertical dot-dashed lines indicate left-to-right: starting point of re-crystallisation, solidus position when at the middle-point of the growth, solidus coordinate at the point of full melting of the feed rod.}
    \label{fig:ModelConcvsMZvsVsolidus}
\end{figure}

The coupled model additionally allows the measured composition profile to be related to the reconstructed molten-zone geometry and local solidification conditions. Figure~\ref{fig:ModelConcvsMZvsVsolidus} compares the calculated composition profiles with the evolution of the equivalent molten-zone height, $\overline{h}_{\mathrm{w}}$, and solidification-interface velocity, $\overline{v}_{\mathrm{s}}$, along the growth direction. Although the imposed translation speed remained constant at 5~mm/h, both the molten-zone height and the actual solidification velocity varied substantially during growth. In particular, changes in $\overline{v}_{\mathrm{s}}$ reflect the continuous evolution of the solid--liquid interface rather than simply the imposed translation of the heating zone. The segregation process therefore takes place under continuously changing local solidification conditions, even though the externally imposed growth parameters remain constant.

\section{Discussion and outlook}
\label{sec:DisOut}

In the present sapphire-confined geometry, the key process variables controlling segregation, including the molten-zone length, the solid–liquid interface geometry, and the internal temperature distribution cannot be measured with sufficient accuracy in-situ during re-solidification for purely ab-initio calculation. The final chemical profile therefore provides one of the few experimentally accessible records of the intrinsic growth history, allowing the model to serve as a physical reconstruction of otherwise inaccessible thermal and solidification conditions throughout the growth process. In the present work this model is used diagnostically: it identifies the mechanisms responsible for the observed composition gradients and shows that they can be explained by the combined effects of molten-zone evolution, interface motion, and elemental partitioning. An independent experimental check of the reconstructed thermal field, provided by EBSD analysis of the quenched growth shown in Fig.~\ref{fig:EBSD}, delineates a convex solidification front between the grown crystal and quenched melt, demonstrating good agreement with the solidus geometry predicted by the thermal model and providing additional support that the model captures the principal radial and axial heat-flow conditions governing the re-solidification.

For the studied composition ($\sim$Ni$_{50}$Mn$_{28.5}$Ga$_{21.5}$), the alloy is martensitic at room temperature and, in the absence of prior mechanical loading, exhibits a self-accommodated hierarchical twin microstructure, visible in the lower half of Fig.~\ref{fig:EBSD}a. Upon heating above 400~K into the austenitic state (Fig.~\ref{fig:EBSD}b), the cubic parent lattice is recovered, with a $\langle 100\rangle$ crystallographic direction aligned with the rod growth direction. Retention of the seed orientation is particularly advantageous for magnetic-shape-memory applications, where specimens must be cut along specific crystallographic directions. For the relatively small 7~mm ingot diameter used here, alignment of the required crystallographic direction with the growth axis permits substantially longer specimens to be extracted along the rod and simplifies their mechanical sectioning and preparation. Notably, direct contact between the melt and the single-crystal sapphire crucible does not prevent the grown crystal from adopting the seed orientation.

The present FEM model is intentionally simplified to provide a computationally accessible description of the coupled thermal and segregation behaviour during laboratory-scale optical floating-zone growth. Most importantly, fluid flow within the molten zone is not explicitly included, although buoyancy-driven convection, Marangoni flow, and flow associated with motion of the solid--liquid interfaces may all affect solute redistribution and the degree of liquid mixing. This limitation becomes particularly relevant in view of the substantial variations in molten-zone geometry and solidification velocity reconstructed in Fig.~\ref{fig:ModelConcvsMZvsVsolidus}. While changes in molten-zone volume and interface position are accounted for explicitly by the coupled thermal and mass-balance model, the effective partition coefficients, $K^*_{\mathrm{Mn}}$ and $K^*_{\mathrm{Ga}}$, necessarily represent the remaining unresolved segregation physics, including equilibrium partitioning, finite solute redistribution, incomplete melt mixing, and possible dependence on the local solidification rate. Future development should therefore couple the present heat-transfer model to explicit fluid-flow and solute-transport calculations, allowing liquid mixing and the onset of interface destabilisation at higher translation speeds to be treated directly rather than absorbed into effective partition coefficients~\cite{Liu2012,Chen2003,Liu2011}. Further improvements could include composition-dependent thermophysical properties and partition coefficients, as well as a more complete treatment of the final cooling stage, where finite solidification time, residual diffusion, and continued partitioning may smooth the terminal composition gradient.

In its present form, the model provides a method for growth optimisation, enabling the possibility of testing parameters such as relative lamp power, translation speed, crystal diameter, seed crystal and feed-rod lengths, and external heat sinks prior to time- and material-consuming growth experiments. Whilst the present model has been parametrised for the OFZ experiment studied here, the generic heat-input formulation makes it readily adaptable to other crucible-confined zone-melting configurations. With only minor modifications, the model can be used to investigate composition segregation arising from incongruent melting, intentional doping, or impurity redistribution in other inter-metallic systems. To facilitate such adaptation and reuse, the COMSOL model template will be made openly available through Zenodo~\cite{Musiienko2026Zenodo}.

\section{Conclusions}
\label{sec:Conclusions}

Sapphire-confined optical floating-zone growth of Ni--Mn--Ga single crystals was demonstrated and combined with a transient finite-element heat-transfer model coupled to an effective molten-zone mass-balance calculation. The experimentally constrained model reproduces the principal longitudinal segregation trends and reconstructs the evolution of the molten-zone geometry and solidification conditions during growth. Independent comparison with a quenched sample further shows that the calculated solidification-front geometry is consistent with that observed by EBSD, supporting the physical validity of the reconstructed thermal field. The EBSD also confirms adoption of the seed crystal orientation utilising this modified FZ growth technique, despite contact of the melt with the single crystalline sapphire crucible.

We show that a constant crucible translation rate does not correspond to a constant solidification velocity. Both the molten-zone height and local solidification velocity vary substantially during growth as the thermal balance evolves, explaining the major departures from ideal Scheil--Pfann behaviour. By reconstructing these otherwise inaccessible growth conditions, the model provides a practical basis for reducing trial-and-error optimisation, limiting material losses, and improving the efficiency and reproducibility of crystal growth.

\section*{Data availability}
\label{sec:DataAvailability}
Data and COMSOL model template files supporting the findings of this study will be made publicly available on Zenodo on the date of publication at \url{https://doi.org/10.5281/zenodo.23474761}.

\section*{Acknowledgments}
\label{sec:Acknowledgments}
Material preparation, crystal growth, and sample characterisation was carried out in the MGML (\url{http://mgml.eu/}), supported within the Czech Research Infrastructures program (project no. LM2023065). The authors acknowledge the assistance provided by the Ferroic Multifunctionalities project, supported by the Ministry of Education, Youth, and Sports of the Czech Republic. Project No. CZ.02.01.01/00/22\_008/0004591, co-funded by the European Union. We also acknowledge the financial support from project 23-04806S supported by the Czech Science Foundation.

\section*{Declaration of generative AI and AI-assisted technologies in the manuscript preparation process}
\label{sec:DeclarationOfAI}
During the preparation of this work, the authors used ChatGPT (OpenAI) in order to assist with English-language editing, improve the clarity and structure of selected passages, and BibTeX/LaTeX formatting tasks. After using this tool/service, the authors reviewed and edited the content as needed and take full responsibility for the content of the published article.

\appendix
\section{Appendix -- Finite-element model and composition profile fitting}
\label{app:A}
The commercially available COMSOL Multiphysics 6.2 software was used for all finite element modelling. A to-scale model of the floating zone experiment was built in axisymmetrical geometry. Optical heating and radiative and convectional cooling were applied by the boundary inward heat flux conditions. The computational domain representing the phase change material was discretised using a structured square-cell mesh with a fixed element size of 0.1 mm. 

\subsection{Governing equations and assumptions}
\label{subsubsec:GoverningEquations}
The heat-transfer model follows the formulation used in the Heat Transfer Module of \mbox{COMSOL} Multiphysics\textsuperscript{\textregistered}~\cite{COMSOLHeatTransferModule62}. The transient heat transfer in the growth system was described by the conservation of energy equation with heat flux:

\begin{equation}
\rho C_p(T) \frac{\partial T}{\partial t} + \bm{\nabla} \cdot \bm{q} = \bm{Q},
\label{eq:ht-conservation}
\end{equation}

where $\rho$ is the density, $C_p(T)$ is heat capacity, $T$ is the temperature, and $Q$ is a volumetric heat source term.

The heat flux is given by Fourier’s law of conduction:
\begin{equation}
\bm{q} = -k(T) \nabla T,
\label{eq:fourier}
\end{equation}
where $k(T)$ is the thermal conductivity (temperature dependent). Temperature dependences of $C_p(T)$ (linear interpolation fit), $k(T)$ (3-region function interpolation fit), material density near phase transformation and phase transformation parameters are extracted from~\cite{brillo2011thermophysical}:

\begin{equation}
k(T)\;[\mathrm{W\cdot(m\cdot K)^{-1}}]=
\begin{cases}
25\cdot\exp\!\left(-\dfrac{70}{T}\right),
& 473.15 \leq T < 1028.15\;\mathrm{K}, \\
12.29-4.0\cdot10^{-3}\cdot T,
& 1028.15 \leq T < 1405.15\;\mathrm{K}, \\
2.1+8.8\cdot10^{-3}\cdot T,
& 1405.15 \leq T \leq 1873.15\;\mathrm{K}.
\end{cases}
\label{eq:kT}
\end{equation}

\begin{equation}
C_p(T)\;[\mathrm{J\cdot (kg\cdot K)^{-1}}]=
\begin{cases}
420 + 0.1682\,(T-473.15), & T < 1008.15\;\mathrm{K}, \\
510 - 2.4615\,(T-1008.15), & 1008.15 \leq T < 1073.15\;\mathrm{K}, \\
350 + 0.15625\,(T-1073.15), & 1073.15 \leq T < 1405.15\;\mathrm{K}, \\
800, & T \geq 1405.15\;\mathrm{K}.
\end{cases}
\label{eq:Cp}
\end{equation}

Radiative exchange between heated surfaces and protective atmosphere, was accounted for using the surface-to-ambient radiation interface in COMSOL Heat Transfer module:
\begin{equation}
-\mathbf{n \cdot q}= \varepsilon \sigma (T^4_{\mathrm{amb}} - T^4),
\label{eq:radiativeflux}
\end{equation}
where $\varepsilon=0.2$ is the surface emissivity, $\sigma$ is Stefan-Boltzmann constant and $T_{\mathrm{amb}}$ is the ambient temperature. A constant effective hemispherical emissivity of $\varepsilon = 0.2$ was assumed for the alloy in both the solid and liquid states. Direct emissivity data for Ni--Mn--Ga near the melting range were not available, so this value was chosen as a reasonable estimate for a clean metallic surface. As a reference, tabulated data for unoxidized nickel give emissivity values close to $\varepsilon \approx 0.19$ at elevated temperature, whereas oxidized nickel surfaces show substantially higher values. The use of a low emissivity is also supported by the experimental observation: during processing, the molten zone could be viewed through the transparent sapphire crucible and appeared glossy, indicating a smooth, reflective surface rather than an oxidized or rough one. Therefore, $\varepsilon = 0.2$ was treated as an effective emissivity for the present thermal model rather than as a directly measured material constant. Convective cooling was described as a constant heat flux rate of $h_{\mathrm{conv}}$: 
\begin{equation}
-\mathbf{n \cdot q}= h_{\mathrm{conv}} (T_{\mathrm{amb}} - T)
\label{eq:convectiveflux}
\end{equation}
Boundary heat influx from the furnace's optical heater was described by an integrally normalised Gaussian profile with standard deviation $\omega/2$. 
To account for the translation of the heating zone, the profile was evaluated in a coordinate system moving with the focal point, following the conditional boundary implementation described by S{\"o}nnerlind~\cite{Sonnerlind2016ConditionalBC}. The imposed heat flux was defined as
\begin{equation}
P(z,t) = \frac{P_{\mathrm{axisym}}}{\omega \sqrt{\pi/2}}\cdot \exp\left[-2\left(\frac{z-z_f(t)}{\omega}\right)^2\right]\;,\;\;z_f(t) = z_{\mathrm{init}}+v_{\mathrm{vert}}\cdot t
\label{eq:powerintegral}
\end{equation}

Phase change is modelled using an apparent heat capacity method, where the latent heat of transformation is incorporated into the energy balance through a temperature-dependent effective heat capacity. In the governing heat Eq.~(\ref{eq:ht-conservation}) heat capacity $C_p(T)$ is expressed as effective heat capacity $C_p^{\mathrm{eff}}(T)$ accounting for latent heat $L$ released or absorbed during phase change. It is defined using a smoothed Heaviside function $H(T)$ as
\begin{equation}
C_p^{\mathrm{eff}}(T) = C_p(T) + L \frac{dH(T)}{dT}.
\label{eq:cp-effective}
\end{equation}

The Heaviside function represents the phase transition over a finite temperature interval $[T_m - \Delta T_{m}/2,\, T_m + \Delta T_{m}/2]$, where $T_m$ is the phase change temperature and $\Delta T_{m}$ is a smoothing parameter, referenced to as \textit{mushy zone}:
\begin{equation}
H(T) =
\begin{cases}
0, & T < T_m - \frac{\Delta T_{m}}{2}, \\
\text{smoothed step function}, & T_m - \frac{\Delta T_{m}}{2} \le T \le T_m + \frac{\Delta T_{m}}{2}, \\
1, & T > T_m + \frac{\Delta T_{m}}{2}.
\end{cases}
\label{eq:heaviside}
\end{equation}
This formulation ensures numerical stability by avoiding discontinuities while approximating the latent heat effect as a distributed contribution to the heat capacity over the transition interval.

\subsection{Modelling process in detail}
\label{subsec:ModellingProcess}
\begin{figure}[pos=htbp]
    \centering
    \includegraphics[width=0.45\linewidth]{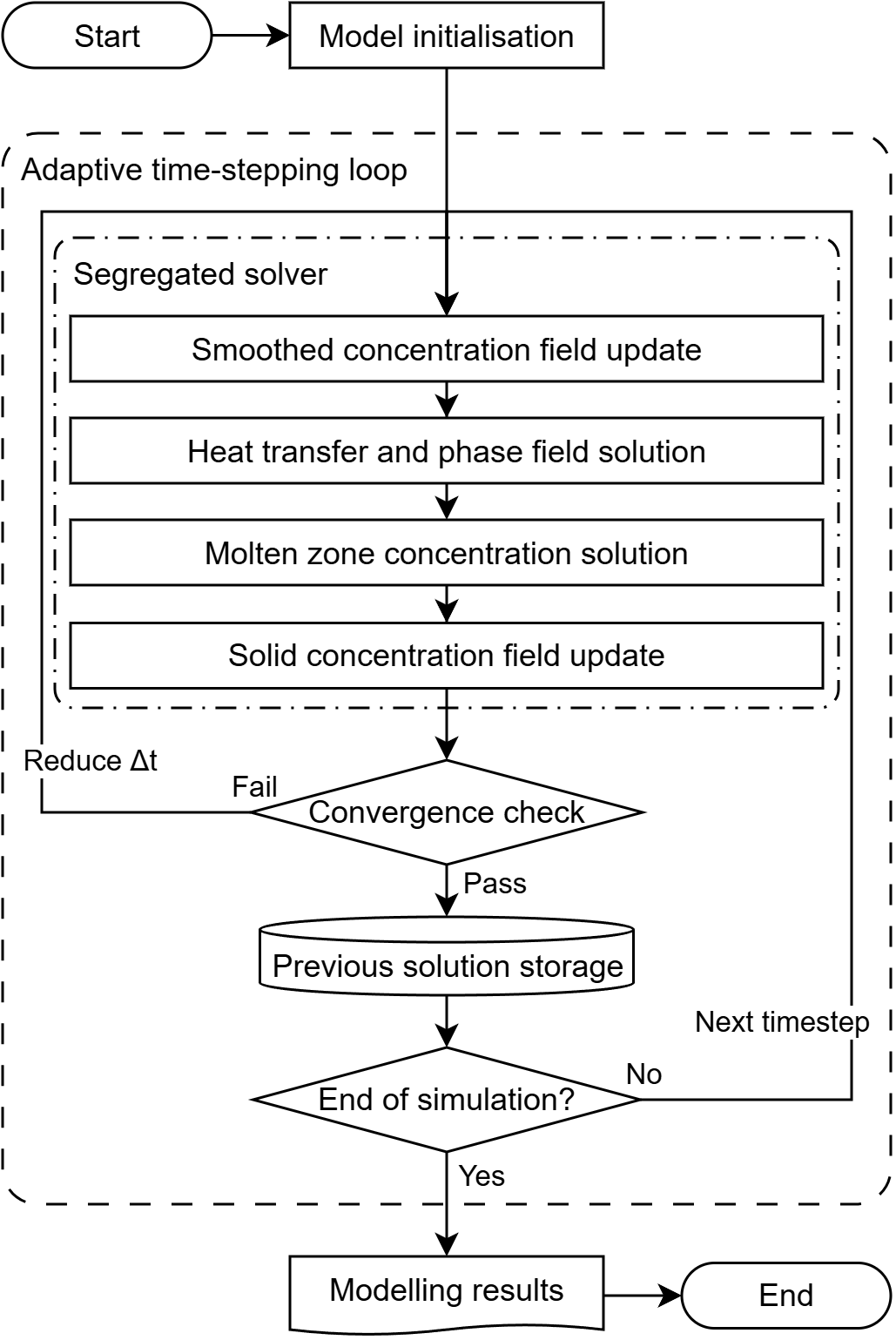}
    \caption{A flow chart describing the computation process and  coupling between the different computation steps.}
    \label{fig:ModelChart}
\end{figure}

A flow-chart of the model calculation process is illustrated in Figure~\ref{fig:ModelChart}. After the initialisation step the model works as follows. At each adaptive time-step, the smoothed composition field is updated for visualisation, followed by heat-transfer and phase-field solution, followed by the computation of ordinary differential equations defining elemental concentrations in the melt and domain state variables for concentrations in fully solidified material. The changes to the concentrations are driven by the phase state time derivatives. The solved state, including thermal, phase and composition fields, then acts as the starting state for the next calculation step and is referenced to as the previous solution.

\begin{figure}[pos=htbp]
    \centering
    \includegraphics[width=0.5\linewidth]{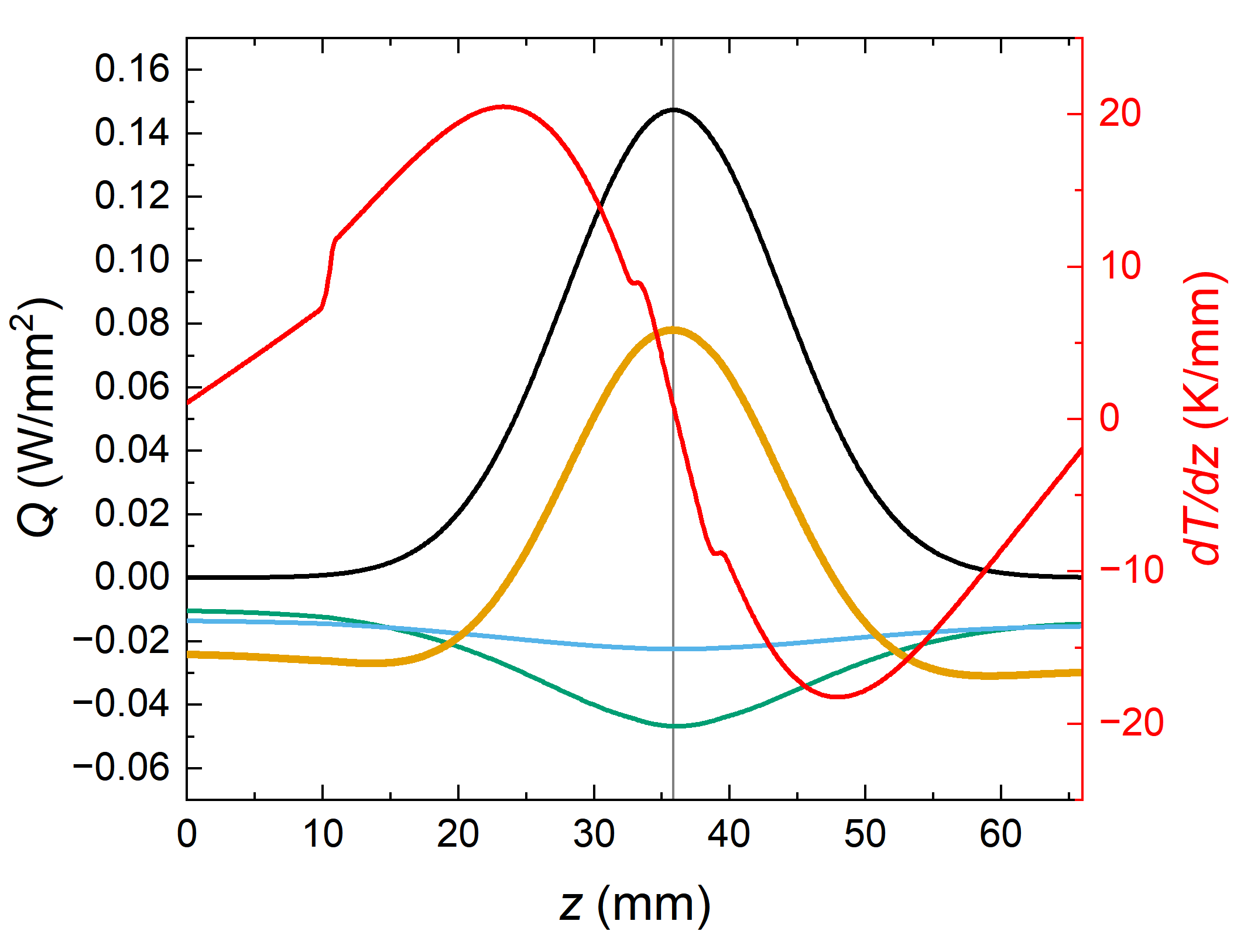}
    \caption{Inward heat fluxes at the heated boundary of the modelled system: deposited boundary heat influx $P(x)$, black curve; "convective" heat influx $h_{\mathrm{conv}} \cdot (T_{\mathrm{amb}}-T)$, cyan curve; radiative heat influx $\varepsilon \sigma (T^4_{\mathrm{amb}} - T^4)$, green curve; effective heat influx, orange curve; temperature gradient $dT/dz$, red curve.}
    \label{fig:HeatFluxes}
\end{figure}

During the simulated experiment, the heating process and translation of the incident heat flux follow that used in a typical growth experiment. Figure~\ref{fig:HeatFluxes} shows a snapshot of inward heat fluxes for the outer boundary and compares them with the thermal gradient along the translation axis at $\sim 0.5\cdot t_{growth}$.Two symmetric kinks in the temperature gradient curve near the furnace focus position, indicated by the solid vertical line in Fig.~\ref{fig:HeatFluxes}, correspond to the edges of the mushy zone, where the material undergoes the solid--liquid phase transformation during melting and re-solidification. The additional kink observed farther from the molten region is attributed to the B2--L2\(_1\) transformation. This feature is retained in the model because the temperature-dependent specific heat capacity and heat-transfer coefficients were implemented using segmented approximations of the experimental data, rather than being represented by a single fitted function.

\begin{figure}[pos=htbp]
    \centering
    \includegraphics[width=0.5\linewidth]{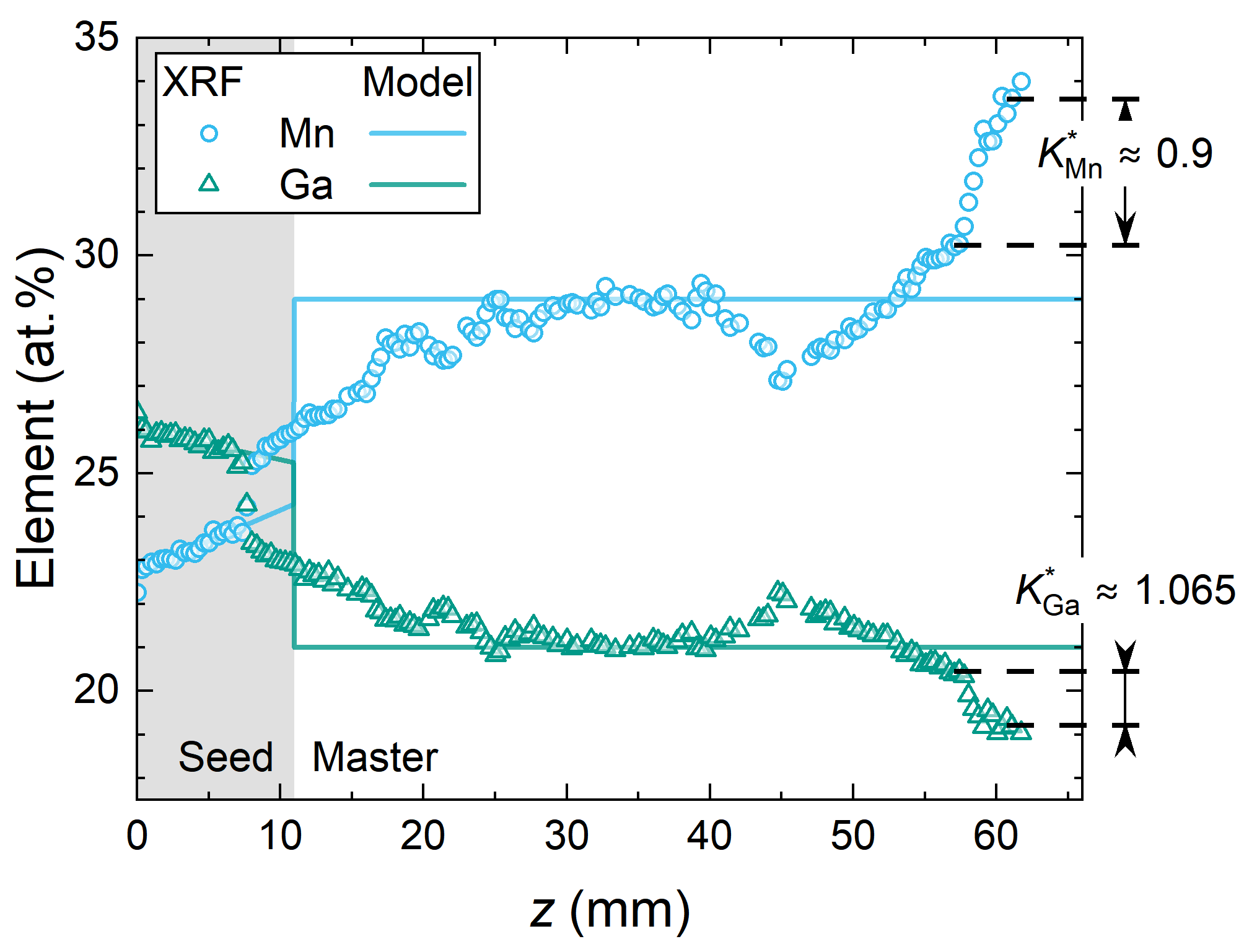}
    \caption{Composition variation of Mn and Ga measured by XRF along the as-grown Ni--Mn--Ga crystal, shown by open symbols. The solid lines show the initial composition distribution used in the model along the axial coordinate, with Mn shown in sky blue and Ga in teal. The filled rectangle marks the seed length. The effective partition coefficients were estimated from the tail of the measured profile, where the rapid composition change occurs.}
    \label{fig:XRFvsModelInitConc}
\end{figure}

Variable model parameters reflecting the growth experiment discussed in this work were set as indicated in Table~\ref{tab:ModelParams}. The initial composition distribution used in the model was re-constructed from X-ray fluorescence measurements of the grown Ni--Mn--Ga crystal. The seeds used for its growth were cut from similarly grown crystals; however, their exact lengths and initial composition gradients were not recorded during the growth procedure. Therefore, for modelling purposes, the seed composition was approximated by a linear gradient fitted to the available XRF data. The resulting initial Mn and Ga concentrations in seed region were expressed as $C^{\mathrm{seed}}_{\mathrm{Mn}} = 0.227+1.415\cdot z/(1\;[\mathrm{m}])$ and $C^{\mathrm{seed}}_{\mathrm{Ga}} = 0.260-0.690\cdot z/(1\;\mathrm{[m]})$ for $0\leq z\leq z^{\mathrm{seed}}_{\mathrm{TopEnd}}$, and $C^{\mathrm{master}}_{\mathrm{Mn}}$ and $C^{\mathrm{master}}_{\mathrm{Ga}}$ for $z>z^{\mathrm{seed}}_{\mathrm{TopEnd}}$, where \(z\) is the axial coordinate along the crystal bar. The corresponding composition profiles are shown in Fig.~\ref{fig:XRFvsModelInitConc}. The initial Ni concentration was assigned as the balance of the ternary composition.

\begin{table}[ht]
    \caption{Variable model parameters used for the present simulation.}
    \label{tab:ModelParams}    
    \centering
    \begin{tabular*}{\tblwidth}{@{}LLL@{}}
        \toprule
        Parameter & Value & Description \\
        \midrule
        {\bf Heat Transfer} && \\[4pt]
        $P_{\mathrm{integral}}$ & 65 [W] & Deposited power from the lamps\\
        $\omega$ & 0.016 [m] & Double standard deviation of the Gaussian Pulse\\
        $v_{\mathrm{vert}}$ & 5 $[\mathrm{mm\cdot h}^{-1}]$ & Vertical translation speed\\
        $z_{\mathrm{init}}$ & 17.8 [mm] & Starting position of the focal centre of the furnace\\
        $t_{\mathrm{heatup}}$ & 1500 [s] & Heat up time\\
        $t_{\mathrm{growth}}$ & $(l_{\mathrm{bar}}-z_{\mathrm{init}}-0.28\;\mathrm{[cm]})/v_{\mathrm{vert}}$ & Total simulation time\\
        $\Delta z_{\mathrm{solver}}$ & 0.1 [mm] & Maximum translation step for solver\\
        $\Delta t_{\mathrm{solver}}$ & $\Delta z_{\mathrm{solver}}/v_{\mathrm{vert}}$ & Maximum time step for solver\\
        $T_{m}$ & 1393.15 [K] & Melting point\\
        $\Delta T_{m}$ & 20 [K] & Temperature interval of phase transformation\\
        $L$ & 223 $[\mathrm{kJ\cdot kg^-1]}$ & Latent heat of melting/solidification\\
        $\rho$ & 6900 $\mathrm{[kg\cdot m^{-3}]}$ & Material density near phase transformation\\
        $l_{\mathrm{bar}}$ & 66 [mm] & Length of the crystal after the growth\\
        $d_{\mathrm{bar}}$ & 7 [mm] & Diameter of the grown crystal\\
        $t_{\mathrm{calculation}}$ & $t_{\mathrm{growth}}+t_{\mathrm{heatup}}$ & Total calculation time\\
        $P_{\mathrm{axisym}}$ & $P_{\mathrm{integral}}\cdot(\pi\cdot d_{\mathrm{bar}})^{-1}$ & Axisymmetrically applied integral power\\
        $h_{\mathrm{conv}}$ & 20 $\mathrm{[W\;m^{-2}\;K^{-1}]}$ & Effective convective heat-transfer coefficient\\
        $\epsilon$ & 0.2 & Surface emissivity\\
        $T_{\mathrm{amb}}$ & 300 $[K]$& Ambient temperature\\
        \midrule
        {\bf Composition calculation} && \\[4pt]
        $C^{\mathrm{master}}_{Mn}$ & 0.290 & Mn atomic fraction in master alloy\\
        $C^{\mathrm{master}}_{Ga}$ & 0.21 & Ga atomic fraction in master alloy\\
        $C^{\mathrm{master}}_{Ni}$ & $1-C^{\mathrm{master}}_{\mathrm{Ga}}-C^{\mathrm{master}}_{\mathrm{Mn}}$ & Ni atomic fraction in master alloy\\
        $z^{\mathrm{seed}}_{\mathrm{TopEnd}}$ & 10.5 [mm] & Coordinate of the top end of the seed\\
        $K^*_{\mathrm{Mn}}$ & 0.900 & Effective Mn partition coefficient\\
        $K^*_{\mathrm{Ga}}$ & 1.065 & Effective Ga partition coefficient\\
        \bottomrule
    \end{tabular*}
\end{table}

The nominal seed length was approximately 1 cm, although variations of up to 1 mm are possible due to the manual cutting procedure. For this reason, the seed top-end position, $z^{\mathrm{seed}}_{\mathrm{TopEnd}}$, and the initial position of the molten zone, $z_{\mathrm{init}}$, were treated as empirical parameters. They were adjusted within the expected uncertainty range in order to reproduce the composition gradient at the beginning of the re-solidification process.

The model tracks Mn and Ga concentrations explicitly, while Ni is calculated as the remaining fraction. This choice was made because Mn and Ga exhibit the largest relative changes in composition in the studied ternary system. Effective partition coefficients, $K^*_{\mathrm{Mn}}$ and $K^*_{\mathrm{Ga}}$, were estimated from the tail region of the re-solidified bar. They were obtained as ratios of average compositions measured immediately before and after the rapid composition change observed near the end of the XRF scan, as indicated in Fig.~\ref{fig:XRFvsModelInitConc}. The total modelled bar length, \(l_{\mathrm{bar}}\), was chosen to cover the experimentally analysed region and the thermally affected part of the sample. 

\subsection{Global variables}
\label{subsec:GlobalVariables}
Positive phase change to liquid and to solid in the node during a time step $\Delta t$:
$$\zeta_L=\Phi_L\cdot\Delta t = 0.5\cdot(\vert\frac{\partial\theta_2}{\partial t}\vert+\frac{\partial\theta_2}{\partial t})\cdot\Delta t $$
$$\zeta_S = \Phi_S\cdot\Delta t = 0.5\cdot(\vert\frac{\partial\theta_1}{\partial t}\vert+\frac{\partial\theta_1}{\partial t})\cdot\Delta t $$

Volumes of material added to and removed from the molten zone during the calculation step $\Delta t$:
$$V^{S \rightarrow L} = \int_V\zeta_Ldv, \;\;\; V^{L \rightarrow S} = \int_V\zeta_Sdv$$

Volume fractions of Mn and Ga added to the MZ during the calculation step $\Delta t$:
$$V^{S \rightarrow L}_{\mathrm{Mn}} = \int_V c^S_{\mathrm{Mn}}\zeta_Ldv,\;\;\;V^{S \rightarrow L}_{\mathrm{Ga}} = \int_V c^S_{\mathrm{Ga}}\zeta_Ldv$$

Concentrations of Mn and Ga in the material solidified during the current calculation step:
$$c^{L \rightarrow S}_{\mathrm{Mn}} = c^{L}_{\mathrm{Mn}} \cdot K^*_{\mathrm{Mn}},\;\;\;c^{L \rightarrow S}_{\mathrm{Ga}} = c^{L}_{\mathrm{Ga}} \cdot K^*_{\mathrm{Ga}}.$$

Volume of the molten zone:
$$V_L = \int_V \theta_2dv$$

Weighted average molten-zone height:
$$\overline{h}_{\mathrm{w}} = \frac{4V_L}{\pi d_{\mathrm{bar}}^2}$$

Weighted average coordinate of the solidification interface:
$$\overline{z}_s=\frac{\int_V(\vert\frac{\partial\theta_1}{\partial t}\vert+\frac{\partial\theta_1}{\partial t})zdv}{\int_V(\vert\frac{\partial\theta_1}{\partial t}\vert+\frac{\partial\theta_1}{\partial t})dv}$$

Velocity of the solidification interface:
$$v_s=\frac{dz_s}{dt}$$

\subsection{Domain ODEs for Previous Solution step}
\label{subsec:DomainODEs}

Maximum temperature record:
$$T_{\mathrm{max}}-nojac(if(T>T_{\mathrm{max},\xi-1},\;T,\;T_{\mathrm{max},\xi-1}))=0,\; T_{\mathrm{max},0}=T_{\mathrm{amb}}$$

Previous phase state:
$$\theta_{1,\xi-1}-nojac(\theta_1)=0,\; \theta_{1,\xi-1,0}=1$$

\subsection{Global ODEs}
\label{subsec:GlobalODEs}
Global ordinary differential equations are used to calculate the change of concentrations in the molten zone:
$$\begin{cases} \frac{V_L\cdot c^L_{\mathrm{Mn}, \xi-1}+V^{S\rightarrow L}_{\mathrm{Mn}}-V^{L\rightarrow S}\cdot c^{L\rightarrow S}_{\mathrm{Mn}}}{V_L+V^{S\rightarrow L}-V^{L\rightarrow S}}-c^{L}_{\mathrm{Mn}}=0, & \mbox{if } V_L>0 \\ c^{L}_{\mathrm{Mn}}=0, & \mbox{if } V_L=0 \end{cases}$$

$$\begin{cases} \frac{V_L\cdot c^L_{\mathrm{Ga}, \xi-1}+V^{S\rightarrow L}_{\mathrm{Ga}}-V^{L\rightarrow S}\cdot c^{L\rightarrow S}_{\mathrm{Ga}}}{V_L+V^{S\rightarrow L}-V^{L\rightarrow S}}-c^{L}_{\mathrm{Ga}}=0, & \mbox{if } V_L>0 \\ c^{L}_{\mathrm{Ga}}=0, & \mbox{if } V_L=0 \end{cases}$$

$$\begin{cases} 1-c^L_{\mathrm{Mn}}-c^L_{\mathrm{Ga}}-c^L_{\mathrm{Ni}}=0, & \mbox{if } V_L>0 \\ c^{L}_{\mathrm{Ni}}=0, & \mbox{if } V_L=0 \end{cases}$$

Previous concentrations in the molten zone for the next solution step are calculated as follows, they are calculated at the last segregated step of calculation loop and are stored as "$\xi-1$" for the step "$\xi$" they will be used in:
$$c^L_{\mathrm{Mn},\xi-1} - nojac(c^L_{\mathrm{Mn}}) = 0,\;\;\; c^L_{\mathrm{Ga},\xi-1} - nojac(c^L_{\mathrm{Ga}}) = 0$$

\subsection{Domain state variables}
\label{subsec:StateVariables}
Following variables are used to store the concentrations of Mn, Ga and Ni in the ingot with smooth transition through the mushy zone.
$$c_{\mathrm{Mn}}=c^S_{\mathrm{Mn}}\cdot \theta_1 + c^L_{\mathrm{Mn}}\cdot \theta_2,\;\;\; c_{\mathrm{Ga}}=c^S_{\mathrm{Ga}}\cdot \theta_1 + c^L_{\mathrm{Ga}}\cdot \theta_2, \;\;\; c_{\mathrm{Ni}}=c^S_{\mathrm{Ni}}\cdot \theta_1 + c^L_{\mathrm{Ni}}\cdot \theta_2$$

The changed concentrations of solid fraction after re-solidification are stored in following variables. 
$$c^S_{\mathrm{Mn}} = \begin{cases} \frac{c^S_{\mathrm{Mn},\xi-1}\cdot\theta_{1,\xi-1}+ c^{L\rightarrow S}_{\mathrm{Mn}} \cdot 0.5\cdot(\vert\frac{\partial\theta_2}{\partial t}\vert-\frac{\partial\theta_2}{\partial t})\cdot \Delta t}{\theta_{1,\xi-1}+0.5\cdot(\vert\frac{\partial\theta_2}{\partial t}\vert-\frac{\partial\theta_2}{\partial t})\cdot \Delta t}, & \mbox{if } \frac{\partial\theta_1}{\partial t}>0 \\ c^{S}_{\mathrm{Mn},\xi-1}, & \mbox{if } \frac{\partial\theta_1}{\partial t}\leqslant0 \end{cases}$$

$$c^S_{\mathrm{Ga}} = \begin{cases} \frac{c^S_{\mathrm{Ga},\xi-1}\cdot\theta_{1,\xi-1}+ c^{L\rightarrow S}_{\mathrm{Ga}} \cdot 0.5\cdot(\vert\frac{\partial\theta_2}{\partial t}\vert-\frac{\partial\theta_2}{\partial t})\cdot \Delta t}{\theta_{1,\xi-1}+0.5\cdot(\vert\frac{\partial\theta_2}{\partial t}\vert-\frac{\partial\theta_2}{\partial t})\cdot \Delta t}, & \mbox{if } \frac{\partial\theta_1}{\partial t}>0 \\ c^{S}_{\mathrm{Ga},\xi-1}, & \mbox{if } \frac{\partial\theta_1}{\partial t}\leqslant0 \end{cases}$$

$$c^S_{\mathrm{Ni}} = 1-c^S_{\mathrm{Mn}}-c^S_{\mathrm{Ga}}$$

Initial values are defined by the distribution of concentrations prior to re-solidification, see Fig.~\ref{fig:XRFvsModelInitConc}.

\subsection{Case-specific identification of effective model parameters}
\label{subsec:AdjustingParameters}
\begin{figure}[pos=htbp]
    \centering
    \includegraphics[width=0.5\linewidth]{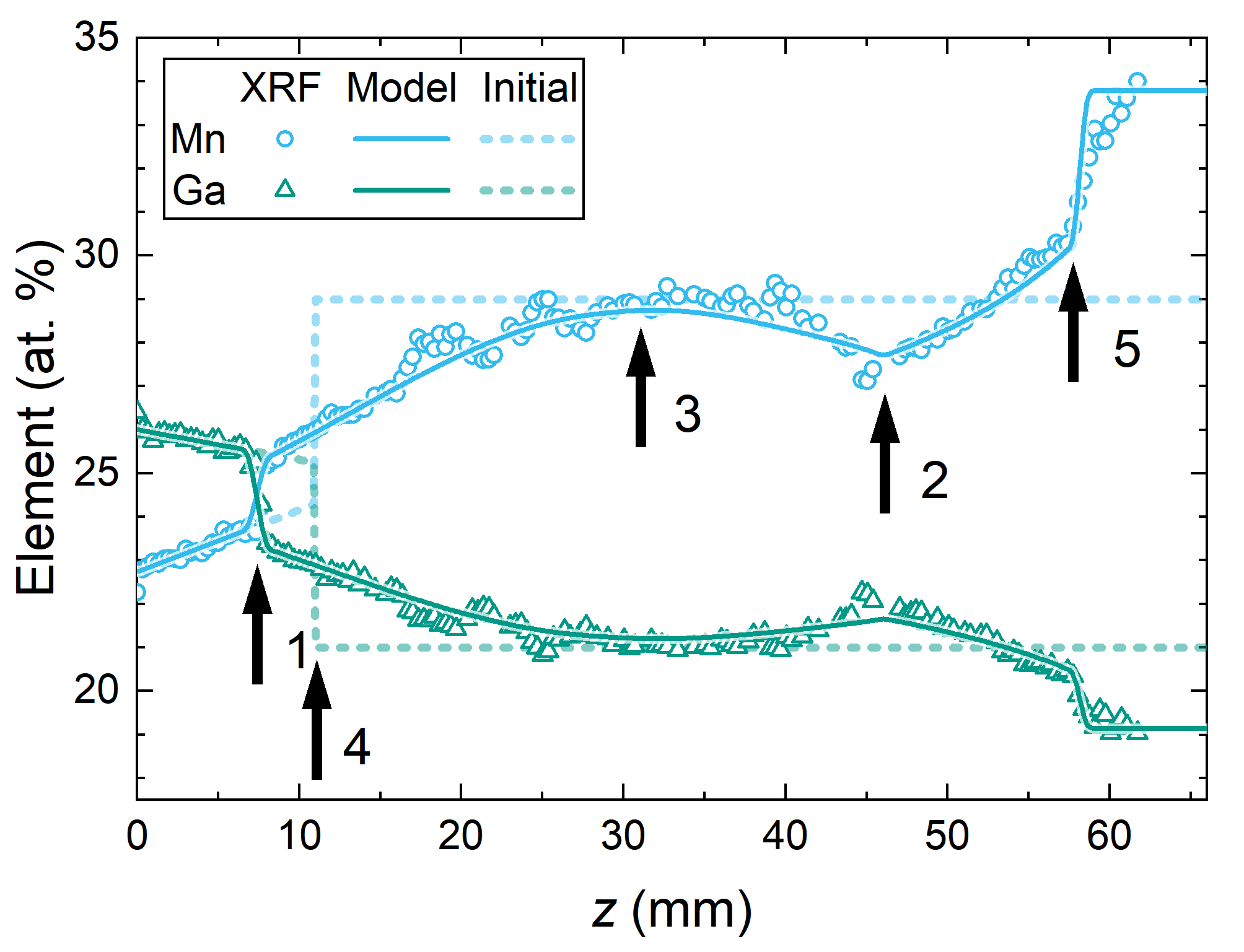}
    \caption{Inverse identification of effective model parameters from the measured Mn and Ga composition profiles. Open symbols show the measured composition along the growth direction for Mn (circles) and Ga (triangles). Solid lines in the foreground represent the modelled composition, solid semitransparent lines in the background show the initial values of $c^S_{\mathrm{Mn}}$ (cyan) and $c^S_{\mathrm{Ga}}$ (green) used in the model.}
    \label{fig:XRF_ModelConcvsMZ}
\end{figure}

To identify a physically reasonable parameter set reproducing the experimental composition profile, a rectangular mesh with an element size of $5 \cdot 10^{-4}$ m was used. Arrows in Fig.~\ref{fig:XRF_ModelConcvsMZ} indicate the focus areas of the XRF data that the model was adjusted to: arrow 1 indicates where the re-solidification started; arrow 2 indicates the bottom mushy zone coordinate at the moment the liquification interface disappears, i.e. when the molten zone reaches the top of the feed rod; arrow 3 indicates the position where $\frac{dV_L}{dt}$ changes sign; arrow 4 - top of the original seed coordinate prior to melting; arrow 5 - the bottom of mushy zone at the moment when lamps were switched off. The calculation is stopped soon after the bottom of mushy zone reaches coordinate (5). The manual adjustment process followed the steps: (i) $l_{\mathrm{bar}}$ was initially adjusted to approximately reproduce the position of feature (3); (ii) total power $P_{\mathrm{integral}}$ and $l_{\mathrm{bar}}$ were then adjusted to fit the position of kink (2) and the magnitude of the bulge (3); (iii) $z_{\mathrm{init}}$ was adjusted to fit the coordinate of composition jump (1); (iv) $z^{\mathrm{seed}}_{\mathrm{TopEnd}}$ and master composition $C^{\mathrm{master}}_{\mathrm{Mn}}$ and $C^{\mathrm{master}}_{\mathrm{Ga}}$ were adjusted for the calculated composition to fit the experimental data. 
Despite reproducing the main features of the experimental profile, a small mismatch remains around position 3. This may arise from the simplified treatment of the total bar length, which in reality changes dynamically as the slightly undersized feed bar melts and fills the sapphire crucible. The calculated composition gradient beyond position 5 is also sharper than expected experimentally, since we assume near-instantaneous freezing of the remaining molten zone and discard calculation data generated by the model beyond the moment the mushy zone surpasses coordinate (5). Allowing finite final-stage solidification, residual diffusion, and continued partitioning would be expected to smooth this terminal composition change.

\printcredits

\bibliographystyle{model1-num-names}

\bibliography{References}

\end{document}